\documentclass[fleqn,usenatbib]{mnras}

\usepackage{newtxtext,newtxmath}

\usepackage[T1]{fontenc}
\usepackage{ae,aecompl}

\usepackage{graphicx}	
\usepackage{amsmath}	
\usepackage{amssymb}	
\usepackage{xspace}
\usepackage{amsfonts,textcomp}
\usepackage[usenames]{color}
\usepackage{times}
\usepackage{hyperref}
\usepackage{ulem}
\usepackage{enumitem}
\usepackage[compatibility=false]{caption}
\usepackage{subcaption}
\usepackage{multirow}
\usepackage{tabulary}
\usepackage[para]{threeparttable}
\usepackage{array,booktabs,longtable,tabularx}
\newcolumntype{L}{>{\raggedright\arraybackslash}X}
\usepackage{ltablex}

\renewlist{tablenotes}{enumerate}{1}
\makeatletter
\setlist[tablenotes]{label=\tnote{\alph*},ref=\alph*,itemsep=\z@,topsep=\z@skip,partopsep=\z@skip,parsep=\z@,itemindent=\z@,labelindent=\tabcolsep,labelsep=.2em,leftmargin=*,align=left,before={\footnotesize}}
\makeatother

\newcommand{\msun}{\mbox{$\rm{M}_\odot$}}
\newcommand{\disperse}{\mbox{{\sc DisPerSE}}\xspace}

\newcommand{\mstar}{\mbox{$M_\star$}\xspace}

\newcommand{\HI}{{\sc H\,i}}

\definecolor{Orange}{rgb}{1.0,0.5,0.15}
\definecolor{Blue}{rgb}{0,0.08,0.65}
\definecolor{Red}{rgb}{0.65,0.08,0.05}
\definecolor{Green}{rgb}{0.15,0.45,0.25}
\definecolor{Pink}{rgb}{1.0,0.05,0.5}

\title[Spin alignment]{SDSS-IV MaNGA: 3D spin alignment of spiral and S0 galaxies}

\author[K. Kraljic et al.]{Katarina Kraljic$^{1,2}$\thanks{E-mail: katarina.kraljic@lam.fr}, Christopher Duckworth$^{3}$, Rita Tojeiro$^{3}$, Shadab Alam$^{2}$, \newauthor Dmitry Bizyaev$^{4,5}$, Anne-Marie Weijmans$^{3}$, Nicholas Fraser Boardman$^{6}$,
Richard R. Lane$^{7,8}$ \\
\\
$^{1}$Institute for Astronomy, University of Edinburgh, Royal Observatory, Blackford Hill, Edinburgh, EH9 3HJ, United Kingdom\\
$^{2}$Aix Marseille Universit\'e, CNRS, CNES, UMR 7326, Laboratoire d'Astrophysique de Marseille, Marseille, France\\
$^{3}$School of Physics and Astronomy, University of St Andrews, North Haugh, St Andrews, KY16 9SS, United Kingdom\\
$^{4}$Apache Point Observatory and New Mexico State University, P.O. Box 59, Sunspot, NM, 88349-0059, USA\\
$^{5}$Sternberg Astronomical Institute, Moscow State University, Moscow, Russia\\
$^{6}$Department of Physics \& Astronomy, University of Utah, Salt Lake City, UT, 84112, USA\\
$^{7}$Instituto de Astronom\'ia y Ciencias Planetarias de Atacama, Universidad de Atacama, Copayapu 485, Copiap\'o, Chile\\
$^{8}$Centro de Investigaci\'on en Astronom\'ia, Universidad Bernardo
O'Higgins, Avenida Viel 1497, Santiago, Chile}

\date{Accepted XXX. Received YYY; in original form ZZZ}

\pubyear{2019}

\begin{document}
\label{firstpage}
\pagerange{\pageref{firstpage}--\pageref{lastpage}}
\maketitle

\begin{abstract}
We investigate the 3D spin alignment of galaxies  with respect to the large-scale filaments using  the MaNGA survey.  
The cosmic web is reconstructed from the Sloan Digital Sky Survey using \disperse and the 3D spins of MaNGA galaxies are estimated using the thin disk approximation with integral field spectroscopy kinematics.   
Late-type spiral galaxies are found to have their spins parallel to the closest filament's axis. The alignment signal is found to be dominated by low-mass spirals. Spins of S0-type galaxies tend to be oriented preferentially in perpendicular direction with respect to the filament's axis. This orthogonal orientation is found to be dominated by S0s that show a notable misalignment between their kinematic components of stellar and ionised gas velocity fields and/or by low mass S0s with lower rotation support compared to their high mass counterparts. 
Qualitatively similar results are obtained when splitting galaxies based on the degree of ordered stellar rotation, such that
galaxies with high spin magnitude have their spin aligned, and those with low spin magnitude in perpendicular direction to the filaments.
In the context of conditional tidal torque theory, these findings suggest that galaxies' spins retain memory of their larger-scale environment. 
In agreement with measurements from hydrodynamical cosmological simulations, the measured signal at low redshift is weak, yet statistically significant.
The dependence of the spin-filament orientation of galaxies on their stellar mass, morphology and kinematics highlights the importance of sample selection to detect the signal.

\end{abstract}



\section{Introduction}

Angular momentum of galaxies is one of the key ingredients to understand their morphological diversity. 
As galaxies are on large scales organised within the network of filaments and walls, it is expected that this large-scale anisotropic environment is, at least partially, driving their morphology.  
Indeed, revisiting the tidal torque theory \citep[TTT;][see also \citeauthor{schafer09} \citeyear{schafer09}, for a review]{Hoyle49,peebles69,doroshkevich70,white84,CatelanTheuns1996,LeePen2000,LeePen2001} in the context of such anisotropic environment, \cite{Codis2015a} explained the relative angular momentum distribution of halos with respect to the filaments and walls of the cosmic web. The misalignment between the tidal and the inertia tensors constrained to the vicinity of filament-type saddle points implies a spin aligned with filaments for low mass halos, and a perpendicular spin orientation for more massive halos, in agreement with findings from cosmological N-body simulations \citep[e.g.][]{AragonCalvo2007,Hahn2007,Codis2012,Trowland2013,WangKang2017,GaneshaiahVeena2018}. 
Recently,
it was found that this transition mass (the critical mass at which the transition of the halo spin orientation occurs) is sensitive to the total neutrino mass \citep{LeeLibeskind2020b} and to the dark energy model \citep{LeeLibeskind2020a}.
The intrinsic spin-shear alignment is therefore potentially powerful complementary probe of massive neutrinos and dark energy.

Galaxies seem to retain a memory of their spin orientation with respect to the cosmic web filaments and walls, as suggested by the results from large-scale cosmological hydrodynamical simulations \citep[][]{Dubois2014,Codis2018,Wang2018,GaneshaiahVeena2019,Kraljic2020a}. The mass dependence of the spin alignment signal is however debated. 
While some works confirmed the existence of a galaxy spin transition from parallel to perpendicular with respect to the filament's direction \citep{Dubois2014,Codis2018,Kraljic2020a}, and analogously with respect to walls \citep{Codis2018,Kraljic2020a}, others  \citep{GaneshaiahVeena2019,Krolewski2019} found preferential perpendicular orientation with respect to filaments at all masses with no sign of a spin transition. 
A possible interpretation of this lack of detection of a clear transition is the nature of the filaments, with galaxies in thinner filaments having their spins more likely perpendicular to the filament's axis, compared to galaxies of similar mass in thicker filaments \citep{GaneshaiahVeena2019}. This can be in turn understood recalling the multi-scale nature of the problem and the conditional TTT \citep{Codis2015a} predicting larger transition mass for denser, thus thicker, filaments. 
Further support for this interpretation was provided by the findings 
of stronger impact of large scale tides on the galaxy spin orientation in denser filaments \citep[][using filament density as a proxy for the thickness of filaments]{Kraljic2020a}. In addition to the stellar mass, the spin-filament alignment was shown to depend on other internal properties of galaxies. Blue or rotation-supported galaxies were found to dominate the alignment signal at low stellar mass, while red or dispersion-dominated galaxies tend to show a preferential perpendicular alignment \citep{Codis2018,Wang2018,Kraljic2020a}.

The orientation of the galaxies' spin with respect to their large-scale environment has been quite extensively studied also on the observational side. 
When focusing on disc galaxies, some groups find preferentially parallel orientation for spirals \citep[][]{Tempel2013,Tempel2013b}, Scd types \citep{Hirv2017}, or both red and blue galaxies \citep{Zhang2013}. Others report a tendency for a perpendicular orientation for spirals \citep{LeeErdogdu2007,Jones2010,Zhang2015} and Sab galaxies \citep{Hirv2017}, or no evidence for a clear signal \citep{Pahwa2016,Krolewski2019}.
A much better agreement seems to exist for elliptical/S0 galaxies, for which a preferential orthogonal orientation of their spin (or minor axis) with respect to their host filaments is found \citep{Tempel2013,Pahwa2016}, in agreement with 
with results of shape measurements \citep[e.g.][]{OkumuraJing2009,Joachimi2011,Singh2015,Chen2019,Johnston2019}.

More recently, \cite{Welker2020} studied the alignment of galactic spin in projection with respect to the filaments of the cosmic web using the integral field spectroscopic (IFS) survey SAMI. They found a mass-dependence of the signal with low-mass galaxies aligning their spin with their nearest filament while  their higher mass counterparts were found to more likely display an orthogonal orientation.  
Similarly, \cite{BlueBird2020} found a preferential alignment of projected galaxy spins (using the photometric major axis of the stellar disc) with the cosmic filaments when analysing a sample of \HI{} biased late-type low mass galaxies in the  
COSMOS \HI{} Large Extragalactic Survey (CHILES) sample.
On the other hand, \cite{Krolewski2019} found no evidence for alignment in projection between galaxy spins, measured from Mapping Nearby Galaxies at Apache Point Observatory (MaNGA) kinematics and filaments from the SDSS Main Galaxy Sample. There is currently no clear consensus on these various observations. However, these differences are not surprising, given that the signal is expected to be weak and various selection criteria need to be carefully examined.

In this work, we measure the alignment between filaments identified in the SDSS Main Galaxy Sample and galaxy spins measured from MaNGA kinematics in 3D. This is the first measurement of the spin alignment in 3D with the use of IFU spectroscopy kinematics. 
The outline of this paper is as follows. 
In Section~\ref{sec:data}, we briefly describe the data and methods used in this work. Section~\ref{sec:alignment} presents the results on the alignment of spiral and S0-type galaxies. In Section~\ref{sec:conclusion}, we summarise and discuss our findings.

\section{Data and methods}
\label{sec:data}

\subsection{The MaNGA survey}

The MaNGA survey \citep[Mapping Galaxies at Apache Point;][]{bundy2015,wake2017}, one of three programs in the fourth generation of the Sloan Digital Sky Survey \citep[SDSS-IV;][]{blanton2017}, is designed to investigate the internal structure of $\sim$10000 galaxies in the nearby Universe ($0 < z < 0.15$). Detailed kinematics are enabled by integral field unit (IFU) spectroscopy, which uses the 2.5-metre telescope at the Apache Point Observatory \citep{gunn2006} along with the two channel BOSS spectrographs \citep{smee2013} and the MaNGA IFUs \citep{drory2015}. MaNGA provides spatial resolution on kpc scales (2'' diameter fibres) while covering 3600-10300$\mathring{A}$ in wavelength with a spectral resolution of R$\sim$2000. 

MaNGA observations are covered plate by plate, employing a dithered pattern for each galaxy corresponding to one of the 17 fibre-bundles of 5 distinct sizes. Any incomplete data release of MaNGA should therefore be unbiased with respect to IFU sizes and hence a reasonable representation of the final sample scheduled to be complete in 2020. The sizes of the IFU are matched to each galaxy in the Primary sample to give a minimum coverage of 1.5 effective radii ($R_{e}$) \citep[][]{law2015obs}. The Secondary sample covers up-to a minimum of 2.5$R_{e}$ and the Colour-Enhanced supplement up-to 1.5$R_{e}$. All together these three sub-samples produce a galaxy population that is unbiased towards morphology, inclination and colour and provides a near flat distribution in stellar mass, $M_{\ast}$. A full description of the MaNGA observing strategy is given in \citet{law2015obs, yan2016obs}. 

The raw data cubes are calibrated \citep{yan2016spec} and reduced by the Data Reduction Pipeline \citep[DRP;][]{law2016drp}. The fibre flux and inverse variance is extracted from each exposure, which are then wavelength calibrated, flat-fielded and sky subtracted. In this work we use 4633 unique galaxies from the 7th data release of MaNGA (known internally as MaNGA Product Launch 7), which corresponds to data released as part of SDSS-IV data release 15 \citep[][]{sloanDR15}.

\subsection{Stellar kinematics}
\label{subsec:kinematics}

For a complete description of the estimation of stellar kinematics, we direct the reader to the MaNGA Data Analysis Pipeline \citep[DAP;][]{westfall2019} from which our data are taken from. We summarise the key points here. Stellar kinematics are extracted using the Penalised Pixel-Fitting method \citep{cappellari2004, cappellari2017}, which fits the stellar continuum on a spaxel (spectral pixel) by spaxel basis. The line of sight of velocity dispersion is extracted and then absorption-line spectra are fitted using a set 49 clusters of stellar spectra from the MILES stellar library \citep{sanchez2006,falcon2011}. We use spectra that are spatially Voronoi binned to $g$-band \textit{S/N} $\sim$ 10 while excluding any individual spectrum with a $g$-band \textit{S/N} < 1 \citep{cappellari2003}. 

We estimate the global position angle (PA) of the stellar velocity fields using the \texttt{fit\_kinematic\_pa} routine \citep[see Appendix C of][]{krajnovic2006}. We remove background galaxies within the IFU and define PAs in the interval 0-180$^{\circ}$ (east of north).  
To select a clean sample only galaxies with well defined global PAs are considered \citep[see][for more details]{Duckworth2019, Duckworth2020}.
Note that such a requirement naturally excludes gas-poor and slowly rotating elliptical galaxies \citep[the majority of removed galaxies have largely incomplete velocity fields, see][for more details]{Duckworth2019}. However, as the focus of this work is on late-type and lenticular galaxies (as described below), this does not impact our results.

\subsection{NSA catalogue}
The MaNGA survey is targeted from an adapted version of the NASA Sloan Atlas (NSA \texttt{v1\_0\_1}\footnote{\href{https://www.sdss.org/dr15/manga/manga-target-selection/nsa/}{www.sdss.org/dr15/manga}}) which is based on SDSS imaging \citep[][]{blanton2011}. For each galaxy selected from DR15, 
we take the stellar mass found using a K-correction \citep[][]{blanton2007}, the axial ratio (b/a) and photometric position angle from the parent NSA catalogue. These parameters are derived using a set of elliptical annuli (based on the Petrosian radius), which maintain the stability of circularised fits to photometry but without the systematic biases of two-component S\'ersic fits \citep[][]{wake2017}.

\subsection{Morphology}

Owing to the method used to compute the 3D angular momentum of galaxies, we focus in this work on galaxies possessing a disk component, namely late-type (LTGs) and lenticular (S0s) galaxies. 

To classify galaxies into LTGs, we use morphological classifications taken from the second iteration of the citizen science project; GalaxyZoo \citep[GZ;][]{willett2013}. GZ relies on visual inspection from volunteers of the general public, who answer a series of questions to characterise morphological type and identify finer features of each galaxy \citep[see Figure 1 in][for specific questions]{willett2013}.
The final catalogue provides a set of weighted responses from all volunteers which are averaged to provide a probability for each classification (e.g. 'Is the galaxy smooth and rounded with no sign of a disk?'). Classifications in MaNGA are taken from a combination of GZ2 and GZ4 data releases\footnote{\href{https://www.sdss.org/dr15/data_access/value-added-catalogs/?vac_id=manga-morphologies-from-galaxy-zoo}{manga-morphologies-from-galaxy-zoo}}. 

Identification of S0s relies on morphological classifications based on the 
the Deep Learning (DL) scheme introduced in \citet{Dominguez2018}; here we give a quick outline. The algorithm is trained on two visually-based morphological catalogues; citizen science project GZ2 and expert classifications of \citet{nair2010}. Morphology is defined for each MaNGA galaxy using SDSS photometry. The DL algorithm estimates a morphological T-type \citep[e.g. see;][for more information]{nair2010} which is first used to separate each galaxy into ETGs and LTGs. To select lenticulars, all ETGs are assigned a probability of being S0 ($P_{S0}$) based on the presence of disk structure and dominance of the bulge. All DL classifications are finally eye-balled to check for reliability.

The choice of the DL classification for the construction of the sample of lenticulars instead of GZ is motivated by a need of a clean sample of S0s. GZ classification allows for classification of galaxies into a S0-Sa type \citep[from an empirical formula transforming vote fractions; Equation 19 in][]{willett2013}, which is not optimal for the purposes of this study. Visual inspection of galaxies falling in this category confirmed the presence of a large fraction of galaxies with prominent spiral arms, that are absent from the sample of S0s based on the DL scheme. 

On the other hand, GZ classification allows for a selection of highly probable LTGs. Such a clean sample of LTGs is crucial for a detection of a spin-filament alignment signal that is expected to be weak.

\subsection{The Cosmic web}

The filaments of the cosmic web are extracted using the publicly available code \disperse \citep{Sousbie2011,SousbiePichon2011}\footnote{\href{http://www2.iap.fr/users/sousbie/web/html/indexd41d.html}{http://www2.iap.fr/users/sousbie/web/html/indexd41d.html}}, which identifies cosmic web structures with a parameter- and scale-free topologically motivated algorithm. 
For the purposes of this work, \disperse was run with a 3$\sigma$ persistence threshold on the \cite{Tempel2014} SDSS galaxy catalogue \citep[for more details, see e.g.][]{Duckworth2019,Kraljic2020b}.

\subsection{Angular momentum}

To compute the spin of galaxies, we adopt the thin-disk approximation following \cite{LeeErdogdu2007}. Here we summarise the main steps. 

The spin axis of a galaxy in the local spherical coordinate system can be written as
\begin{equation}
\begin{split}
    \hat{L}_r & = \cos i, \\
    \hat{L}_{\theta} & = (1 - \cos^2 i)^{1/2} \sin \rm {PA}, \\
    \hat{L}_{\phi} & = (1 - \cos^2 i)^{1/2} \cos \rm {PA},
\end{split}    
\end{equation}
where PA is the position angle and $i$ is the inclination angle of a galaxy. 
If not stated differently, the position angle PA used throughout this work refers to the kinematic PA as described in Section~\ref{subsec:kinematics}.
The inclination angle $i$ can be computed, following \cite{Haynes1984}, as
\begin{equation}
    \cos^2 i = \frac{(b/a)^2 - p^2}{1 - p^2},
\end{equation}
where $b/a$ is the axis ratio and $p$ the intrinsic flatness parameter that varies with galaxy morphology \citep{Haynes1984}. This parameter accounts for the fact that in practice, the disk of galaxies has finite thickness and the presence of a bulge would have an impact on the $b/a$. The proposed values for $p$ span range from 0.23 for S0 to 0.1 for Scd-Sdm types. We adopted the mean value of 0.158, but choosing extreme values does not change our results.
The value of $i$ is set to $\pi/2$ if $b/a < p$. 

The equatorial Cartesian coordinates of the unit spin vector can then be written as

\begin{equation}
\label{eq:spin}
\begin{split}
    \hat{L}_x & = \hat{L}_r \sin \alpha \cos \beta + \hat{L}_{\theta} \cos \alpha \cos \beta - \hat{L}_{\phi} \sin \beta \\
    \hat{L}_y & = \hat{L}_r \sin \alpha \sin \beta + \hat{L}_{\theta} \cos \alpha \sin \beta + \hat{L}_{\phi} \cos \beta \\
    \hat{L}_z & = \hat{L}_r \cos \alpha - \hat{L}_{\theta} \sin \alpha,
\end{split}    
\end{equation}
where $\alpha = \pi/2 - {\rm DEC}$ and $\beta = {\rm RA}$, with DEC and RA corresponding to declination and right ascension, respectively.
We apply positive sign to $\hat{L}_r$ to all galaxies. Note that this sign ambiguity is expected to decrease the strength of the alignment signal. 
Alternatively, one can take into account the two-fold ambiguity statistically by considering both the clock wise and counter-clock wise rotations \citep{Lee2011}. Such an approach does not alter the conclusions of this paper (see Appendix~\ref{app:method2}).

\section{Spin alignment}
\label{sec:alignment}

To measure the spin alignment, each galaxy is assigned its closest filament's segment, defined by a pair of points providing the local direction of the filament for a given galaxy. The absolute value of cosine of the angle between the spin of the galaxy and its closest filament, $\lvert \cos \gamma \rvert$ is used to assess the alignment with respect to the filamentary network of the cosmic web. Values of $\lvert \cos \gamma \rvert$ close to 1 indicate that galaxy tends to have its spin aligned with the neighbouring filament, while values close to 0 mean that the spin is in the orthogonal direction to the filament's axis.
In order to increase the statistics of the measured signal, each galaxy is assigned two closest filaments' segments. In addition, we consider only galaxies that are not too faraway from the filaments ($\leq$ 13 Mpc) and not too close to the nodes of the cosmic web ($\geq$ 0.5 Mpc)\footnote{The values of these distances cuts are a result of a compromise between a clean and sufficiently large sample. As expected, including galaxies that are at very large distances from the filamentary network or that are in the highest density regions, typically galaxy groups and clusters, decreases the strength of the signal.}. 
To quantify the likelihood whether the measured alignments are consistent with being derived from a random distribution, a Kolmogorov--Smirnov (KS) test was performed on each distribution. To account for any possible observational bias, instead of comparing to a uniform distribution, we construct 2000 realisations of random samples by shuffling galaxy positions, i.e. by randomising the association between galaxies and their closest filaments. 
For indicative purposes, we show also statistical errors computed by bootstrapping
considering 100 random samples drawn from the parent sample with replacement.

\subsection{Spirals}

To construct the sample of spiral or late-type galaxies (LTGs), we consider 
only galaxies with a fraction of debiased classification votes of 0.9, somewhat more conservatively than in \cite{Duckworth2020}. This choice is motivated by our desire to construct a clean sample of LTGs while keeping a reasonable number of galaxies.

\begin{figure}
\centering\includegraphics[width=0.9\columnwidth]{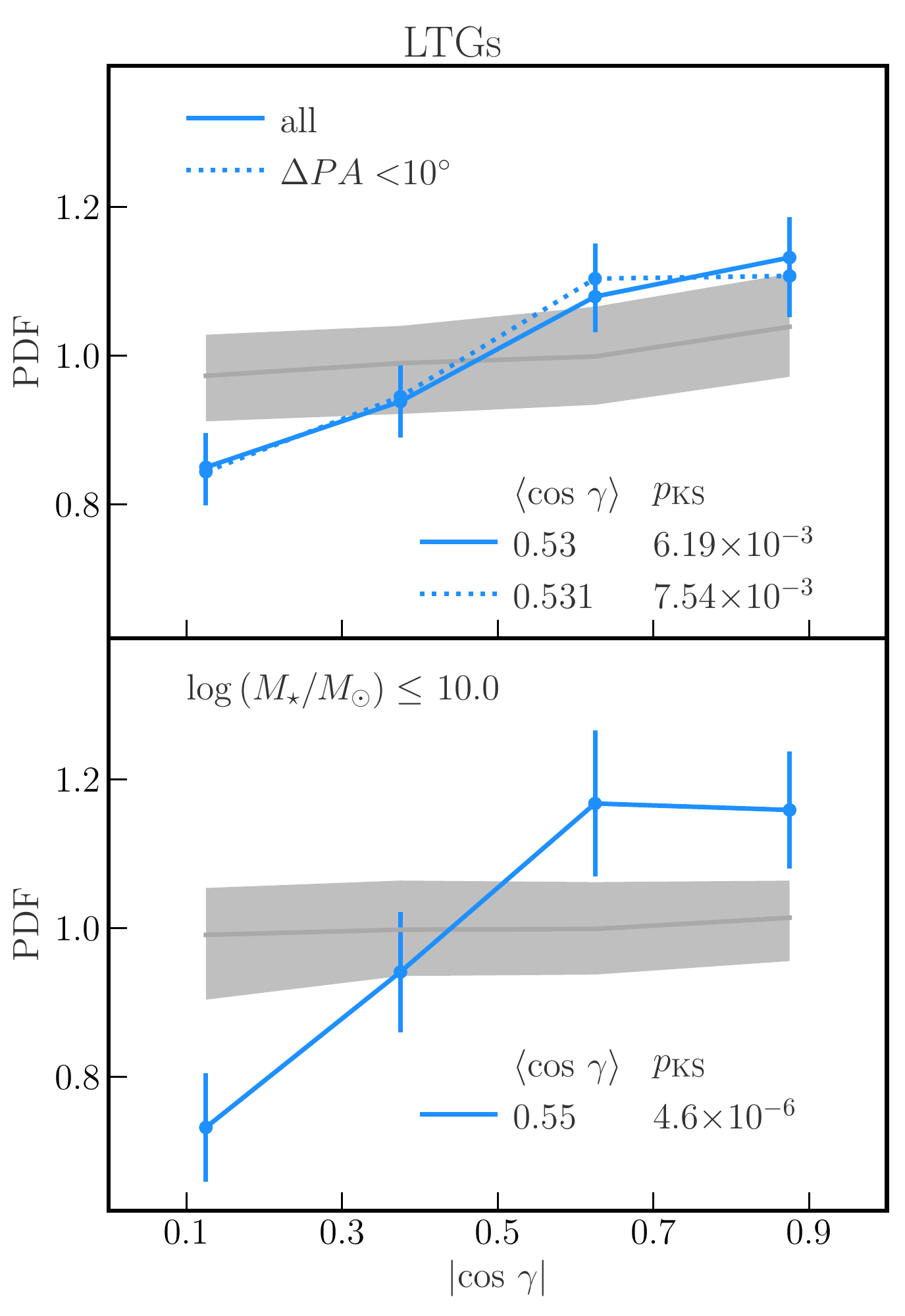}
\caption{Alignment between the filaments and spin of all ({\sl top}), and, low-mass ({\sl bottom}) LTGs. The alignment signal for the kinematically aligned (i.e. $\Delta$PA < 10$^{\circ}$) galaxies only is shown by the dotted line. The error bars correspond to the bootstrap. The solid grey line and shaded area represent the median and 95 per cent confidence limits from 2000 random samples, respectively. The mean alignment angle together with the probability $p_{\rm KS}$ of the KS test are shown in each panel with corresponding symbols. 
LTGs in the probed mass range tend to have their spin parallel to their host filaments. The alignment signal seen for the entire population of LTGs is driven by the low mass sub-sample (\mstar $\leq 10^{10}$ \msun).
}
\label{fig:spin_ltgs}
\end{figure}

Figure~\ref{fig:spin_ltgs} shows the probability distribution function (PDF) of the cosine of the angle between the spin of galaxies and the direction  of their closest filament $\lvert \cos \gamma \rvert$ for the entire sample of LTGs (\textsl{top}) and for low-mass ($\mstar \leq 10^{10} \msun$) LTGs (\textsl{bottom}). 
Spiral galaxies tend to have their spin aligned with the neighbouring filaments. The alignment signal is found to be dominated by low mass spirals. 
We measure the spin of the galaxies from the stellar component (\textsl{solid lines}), however similar results are obtained when considering galaxies with a low offset between the global position angles of stellar and H$\alpha$ velocity fields, $\Delta {\rm PA} < 10^{\circ}$ (\textsl{dotted line}). 
As shown in \cite{Duckworth2019, Duckworth2020}, only a small fraction of LTGs in the MaNGA sample show a significant misalignment.
Interestingly, when using the photometric position angle from the NSA catalogue to compute the spin of galaxies, the measured signal is still statistically significant ($p_{\rm KS}$ < 0.05), but much weaker (see Table~\ref{tab:summary})\footnote{Note that the number of galaxies with the kinematic PA is lower compared to the photometric measurement due to the additional quality cuts (see Section~\ref{subsec:kinematics}). This does not alter our conclusion.}.

\subsection{S0 galaxies}

To select S0 galaxies we consider all ETGs (T-type $\leq$ 0) with $P_{S0}$ > 0.5, in-keeping with \citet{Dominguez2018}.
The selection criterion for S0 galaxies requiring more than 50 per cent of debiased classification votes is somewhat less conservative compared to the selection of LTGs \citep[and also compared to e.g.][]{Duckworth2019}, and motivated by low number statistics of the sample.\footnote{We note however that restricting this condition to over 70 per cent yields qualitatively similar results, albeit with reduced statistical significance.}  

\begin{figure}
\centering\includegraphics[width=0.9\columnwidth]{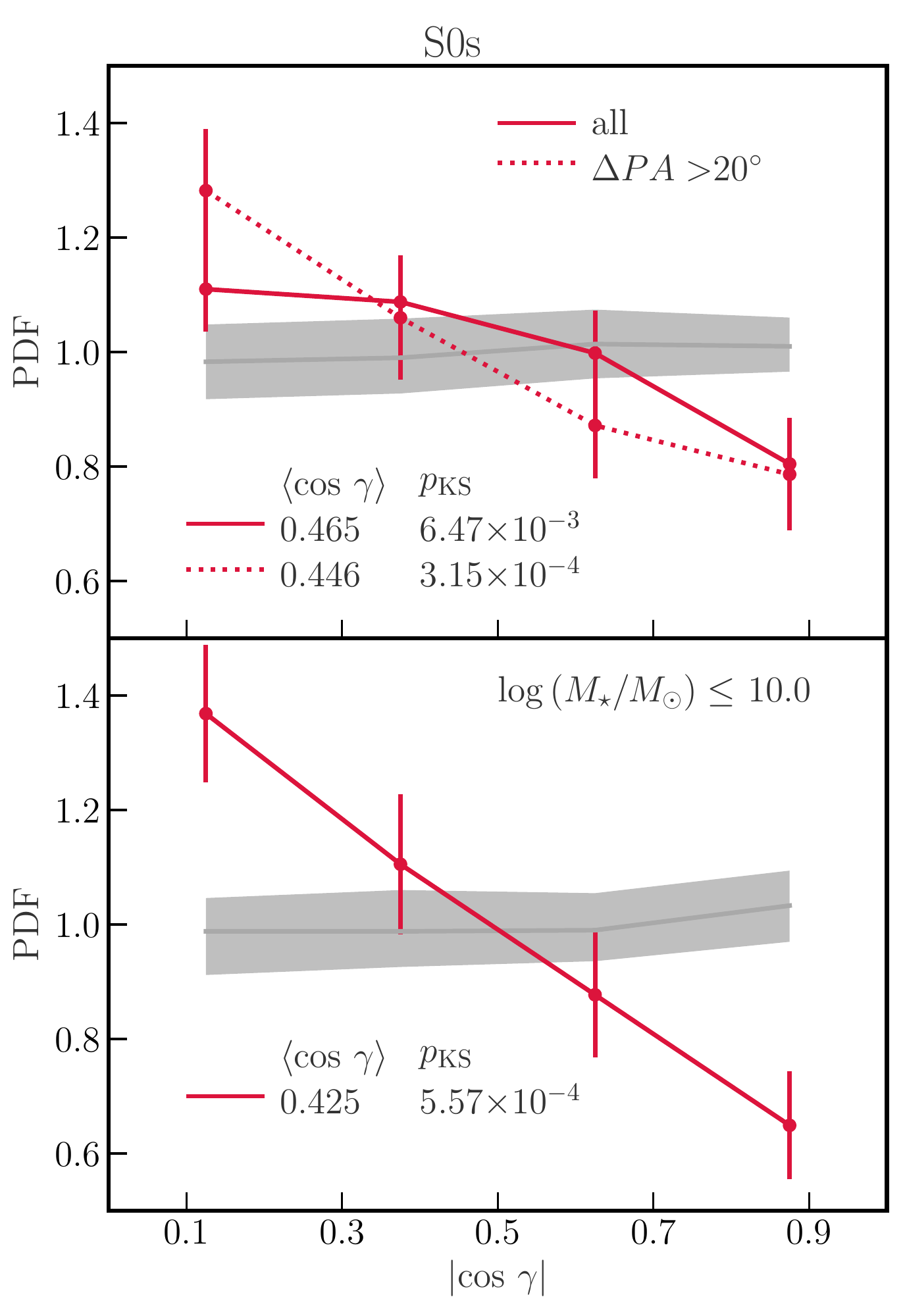}
\caption{Alignment between the filaments and spin of all ({\sl top}) and low-mass ({\sl bottom}) S0s.
The error bars correspond to the bootstrap.  The solid grey line and shaded area represent the median and 95 per cent confidence limits from 2000 random samples, respectively. The mean alignment angle together with the probability $p_{\rm {KS}}$ of the KS test are shown in each panel with corresponding symbols.
S0s in the probed mass range tend to have their spin perpendicular to their host filaments. 
The alignment signal seen for the entire population of S0s is driven by the kinematically misaligned ($\Delta PA > 20^{\circ}$; red dotted line on the top panel) and low mass sub-sample (\mstar $\leq 10^{10}$ \msun). 
}
\label{fig:spin_S0s}
\end{figure}

Figure~\ref{fig:spin_S0s} shows the PDF of the cosine of the angle between the spin of galaxies and the direction  of their closest filament $\lvert \cos \gamma \rvert$ for the entire sample of S0s (\textsl{top}) and for low-mass ($\mstar \leq 10^{10} \msun$) S0s (\textsl{bottom}). 
The spin of S0 galaxies is found to be preferentially orthogonal to the the neighbouring filaments. As for LTGs, the low-mass sub-sample of S0s seem to dominate the signal.
On the other hand, the preferential orthogonal orientation of the spin is found to be driven by kinematically misaligned galaxies, defined as $\Delta {\rm PA} > 20^{\circ}$ (\textsl{dotted line}).
As in the case of LTGs, when using the photometric position angle from the NSA catalogue to compute the spin of galaxies, the measured signal is still statistically significant ($p_{\rm KS}$ < 0.05), but much weaker (see Table~\ref{tab:summary}).

\begin{table}
\centering
\begin{threeparttable}
\caption{Number of galaxies, average $\lvert \cos \gamma \rvert$ and the KS probability $p_{\rm KS}$ that the sample is drawn from a random distribution for various sub-samples of LTGs, S0s, and galaxies with $\lambda_{\ R} > 0.73$ and $\lambda_{\ R} < 0.4$. For LTGs and S0s, we provide also the results for photometric data (shown in parenthesis).
}
\label{tab:summary}
\begin{tabular*}{0.9\columnwidth}{@{\extracolsep{\fill}}lcccc}
\hline
\hline
& selection & $N_{\rm gal}$ & $\left < \cos \gamma \right>$ & $p_{\rm KS}$ \\
\hline
\multirow{7}{*}{LTGs} & \multirow{2}{*}{all} & 611 & 0.53 & 6.2$\times 10^{-3}$\\
& & (852) & (0.515) & (2.3$\times 10^{-2}$)\\
& \multirow{2}{*}{\mstar $\leq 10^{10}$ \msun} & 230 & 0.55 & 4.6$\times 10^{-6}$ \\
& & (377) & (0.521) & (1.9$\times 10^{-2}$)\\
& \multirow{2}{*}{\mstar $> 10^{10}$ \msun} & 381 & 0.517 & 6.4$\times 10^{-1}$ \\
& & (475) & (0.5) & (4.8$\times 10^{-1}$)\\
& $\Delta {\rm PA} < 10^{\circ}$& 556 & 0.53 & 7.5$\times 10^{-3}$ \\
\hline
\multirow{7}{*}{S0s} & \multirow{2}{*}{all} & 269 & 0.465 & 6.5$\times 10^{-3}$\\
& & (363) & (0.482) & (4.5$\times 10^{-2}$)\\
& \multirow{2}{*}{\mstar $\leq 10^{10}$ \msun} & 114 & 0.425 & 5.6$\times 10^{-4}$\\
& & (170) & (0.447) & (2.9$\times 10^{-3}$)\\
& \multirow{2}{*}{\mstar $> 10^{10}$ \msun} & 155 & 0.496 & 5.9$\times 10^{-1}$\\
& & (193) & (0.513) & (5.9$\times 10^{-1}$)\\
& $\Delta {\rm PA} > 20^{\circ}$ & 117 & 0.446 & 3.2$\times 10^{-4}$ \\
\hline
\multirow{4}{*}{$\lambda_{\rm R} > 0.73$ } & all & 131 & 0.554 & 9.0$\times 10^{-3}$\\
& \mstar $\leq 10^{10}$ \msun & 71 & 0.593 & 5.4$\times 10^{-3}$ \\
& \mstar $> 10^{10}$ \msun & 60 & 0.508 & 7.5$\times 10^{-1}$ \\
& $\Delta {\rm PA} < 10^{\circ}$& 130 & 0.557 & 1.0$\times 10^{-2}$ \\
\hline
\multirow{4}{*}{$\lambda_{\rm R} < 0.4$ } & all & 344 & 0.479 & 3.0$\times 10^{-2}$\\
& \mstar $\leq 10^{10}$ \msun & 133 & 0.447 & 6.5$\times 10^{-5}$ \\
& \mstar $> 10^{10}$ \msun & 211 & 0.499 & 3.4$\times 10^{-1}$ \\
& $\Delta {\rm PA} > 20^{\circ}$& 116 & 0.445 & 2.2$\times 10^{-3}$ \\
\hline
\hline
\end{tabular*}
\end{threeparttable}
\end{table}

\section{Discussion and conclusions}
\label{sec:conclusion}

In this work, we present the first measurement of the alignment of galaxy spins with respect to their cosmic filaments with the use of IFS kinematics in 3D. Galaxy spin is reconstructed using the thin-disk approximation applied to galaxies from the MaNGA survey. The filamentary network is reconstructed in 3D from the galaxy distribution in the SDSS survey. Our main results are as follows.

\begin{itemize}
    \item LTGs/spiral galaxies tend to have their spin aligned with the axis of their neighbouring filament. 
    \item The alignment signal for LTGs is dominated by low-mass galaxies, with \mstar $\leq 10^{10}$ \msun.
    \item S0-type galaxies have their spin preferentially in orthogonal direction with respect to their closest filament.
    \item The perpendicular orientation of S0s' spins is dominated by low-mass (\mstar $\leq 10^{10}$ \msun) and/or kinematically misaligned galaxies, with $\Delta {\rm PA} > 20^\circ$.
    \item Qualitatively similar trends are recovered when using photometric data for the spin de-projection instead of IFS kinematics, albeit weaker and with lower statistical significance.
\end{itemize}

Our findings of the preferential spin-filament alignment for spiral galaxies and orthogonal orientation for S0s is in agreement with \cite{Tempel2013} studying the alignment of these two populations of galaxies using the SDSS Main Galaxy sample.
These results are also broadly in agreement with the trends seen in hydrodynamical simulations \citep{Codis2018,Kraljic2020a} showing that at fixed stellar mass the alignment signal is dominated by galaxies with high $v/\sigma$ (ratio of rotation to dispersion dominated velocities, used as a proxy for morphology), while the perpendicular orientation is mostly dominated by galaxies with low values of $v/\sigma$.

\begin{figure*}
\centering\includegraphics[width=0.8\textwidth]{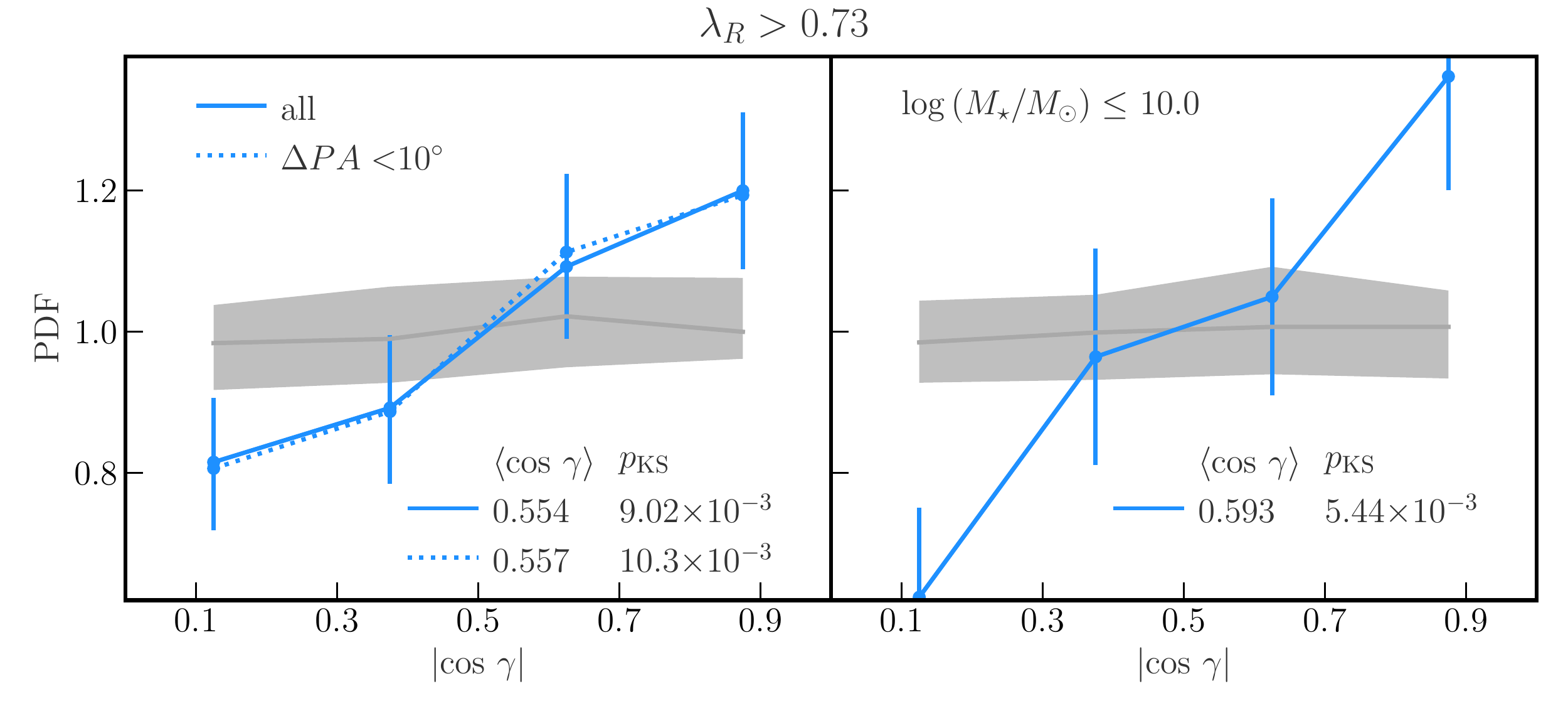}\\
\centering\includegraphics[width=0.8\textwidth]{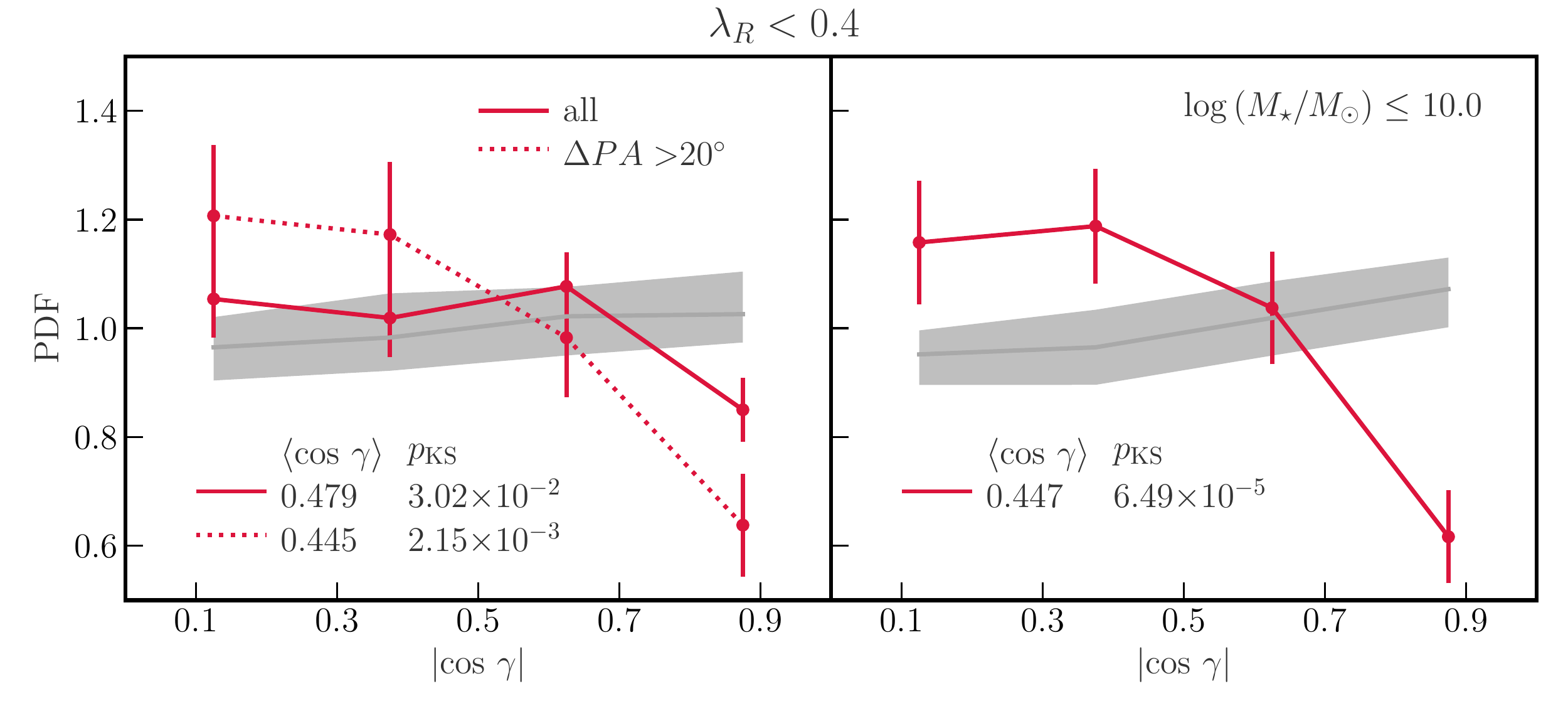}
\caption{{\sl Top:} Alignment between the filaments and spin of all ({\sl left}) and low-mass ({\sl right}) galaxies with high spin magnitude ($\lambda_{\rm R}$ > 0.73). 
{\sl Bottom:} Alignment between the filaments and spin of all ({\sl left}) and low-mass ({\sl right}) galaxies with low spin magnitude ($\lambda_{\rm R}$ < 0.4). 
The error bars correspond to the bootstrap. The solid grey line and shaded area represent the median and 95 per cent confidence limits from 2000 random samples, respectively. The mean alignment angle together with the probability $p_{\rm KS}$ of the KS test are shown in each panel with corresponding symbols.
Galaxies with high spin magnitude in the probed mass range tend to have their spin parallel to their host filaments. The alignment signal seen for the entire population is driven by the low mass sub-sample (\mstar $\leq 10^{10}$ \msun).
Galaxies with low spin magnitude tend to have their spin oriented in perpendicular direction with respect to the filaments. 
The signal seen for the entire population is driven by the the kinematically misaligned ($\Delta PA > 20^{\circ}$; red dotted line on the left panel) and low mass sub-sample (\mstar $\leq 10^{10}$ \msun).
}
\label{fig:spin_rotators}
\end{figure*} 
A decreasing strength of the alignment signal for MaNGA spirals with increasing stellar mass is also in a good agreement with these simulations \citep[and reported stellar mass dependent projected spin flip for SAMI galaxies,][]{Welker2020}. 
Our results for S0-type galaxies appear to be at odds with the predicted stellar mass dependence of the signal.
One would naively expect that massive S0s should dominate the perpendicular spin-filament orientation signal, while the opposite is found. We note however, that a direct comparison with the simulations is not straightforward. This is because the thin-disk approximation used in this work to de-project the spin of galaxies cannot be applied to pure early-type galaxies that are found to dominate the orthogonal spin-filament orientation of massive simulated galaxies. S0-type galaxies are not all slow rotators, they occupy a broad range of the ellipticity ($\epsilon$) -stellar angular momentum estimator \citep[$\lambda_{\rm R}$; see e.g.][]{emsellem2007} parameter space. In addition, kinematically misaligned galaxies tend to reside closer to the slow rotator regime regardless their morphology type \citep[see][their Fig.~8]{Duckworth2020} and to populate lower stellar masses compared with aligned galaxies of the same morphology \citep{Duckworth2019}. 
This is consistent with our finding of low-mass and/or misaligned S0s dominating the orthogonal spin-filament orientation. 
We also find that on average, massive S0s tend to have higher values of $\lambda_{\rm R}$ \citep[see][and their Eq.~2 for details on the computation of the luminosity weighted $\lambda_{\rm R}$ from the IFU data]{Duckworth2020} 
compared to their lower mass counterparts (0.364 $\pm$ 0.017 and 0.323 $\pm$ 0.011, respectively) and compared to kinematically misaligned counterparts (0.24 $\pm$ 0.01). 
A question that naturally arises is whether the parallel vs orthogonal orientation of spin of galaxies with respect to filaments of the cosmic web reflects 
the degree of ordered (stellar) rotation rather than their morphology. To this purpose, we combined all LTGs and S0s, and split the sample based on their $\lambda_{\rm R}$. 
The strongest alignment signal was obtained for galaxies with $\lambda_{\rm R} > 0.73$ and orthogonal signal for galaxies with $\lambda_{\rm R} < 0.4$. 
Galaxies with $\lambda_{\rm R} > 0.73$ tend to be more aligned (higher $\left < \cos \gamma \right>$) than the morphology-based sample of LTGs, however the statistical significance of the signal is somewhat lower ($p_{\rm KS}$ values tend to be larger, see Table~\ref{tab:summary}).
On the other hand, galaxies with $\lambda_{\rm R} < 0.4$ show weaker orthogonal signal (higher $\left < \cos \gamma \right>$) compared to the  morphology-based sample of S0s. 
While this suggests that morphology plays a role in determining the spin orientation with respect to the cosmic web filaments, it is hard at this stage to disentangle kinematics from morphology. Our analysis shows that splitting in either property gives different, but consistent cosmic web alignment signal, however we caution that much larger statistical sample is needed to study the dependence of the spin alignment at fixed stellar mass and morphology for slow and fast rotators.

This work underlines the importance of carefully constructing the sample of galaxies when assessing the spin-filament signal,  given how it depends on stellar mass, morphological type and degree to which their rotation is dominated by dispersion motion. Together with the expectation that at low redshift the alignment is weak \citep[or weaker compared to high redshift, see e.g.][]{Codis2018}, it could also explain the difficulty in detecting the signal in previous studies \citep[e.g.][]{Krolewski2019}. 
It also highlights that the IFS kinematics allows to recover a stronger signal compared to photometric data alone. 
Altogether, this clearly motivates the need for larger IFS surveys such as e.g. the IFS Hector survey \citep{BlandHawthorn2011,Bryant2016}.
These upcoming surveys will be spanning not only large range of stellar masses, but also large-scale environment, allowing to  
extend the existing studies on the spin-filament alignment using MaNGA \cite[this work and that of][]{Krolewski2019} and SAMI IFS surveys \citep{Welker2020}.

\section{Data availability}
The data underlying this article are publicly available and will be shared on reasonable request to the corresponding author.

\section*{Acknowledgements}
The authors thank the anonymous referee for comments that helped improve the clarity of the paper.
We thank St\'ephane Rouberol for the smooth running of the HORIZON Cluster, where some of the post-processing was carried out.
We thank Thierry Sousbie for provision of the {\sc Disperse} code (\href{http://ascl.net/1302.015}{ascl.net/1302.015}). 
KK thanks Christophe Pichon for his exceptional support, enthusiasm and exacting attention to details that have been a great source of inspiration. 
KK also warmly thanks Jounghun Lee for enlightening discussions and helpful feedback.

Funding for the Sloan Digital Sky Survey IV has been provided by the 
Alfred P. Sloan Foundation, the U.S. Department of Energy Office of 
Science, and the Participating Institutions. 

SDSS-IV acknowledges support and resources from the Center for High 
Performance Computing  at the University of Utah. The SDSS website is www.sdss.org.

SDSS-IV is managed by the Astrophysical Research Consortium 
for the Participating Institutions of the SDSS Collaboration including 
the Brazilian Participation Group, the Carnegie Institution for Science, 
Carnegie Mellon University, Center for Astrophysics | Harvard \& 
Smithsonian, the Chilean Participation Group, the French Participation Group, Instituto de Astrof\'isica de Canarias, The Johns Hopkins 
University, Kavli Institute for the Physics and Mathematics of the 
Universe (IPMU) / University of Tokyo, the Korean Participation Group, 
Lawrence Berkeley National Laboratory, Leibniz Institut f\"ur Astrophysik 
Potsdam (AIP),  Max-Planck-Institut f\"ur Astronomie (MPIA Heidelberg), 
Max-Planck-Institut f\"ur Astrophysik (MPA Garching), 
Max-Planck-Institut f\"ur Extraterrestrische Physik (MPE), National Astronomical Observatories of China, New Mexico State University, New York University, University of Notre Dame, Observat\'ario 
Nacional / MCTI, The Ohio State University, Pennsylvania State University, Shanghai Astronomical Observatory, United Kingdom Participation Group, Universidad Nacional Aut\'onoma 
de M\'exico, University of Arizona, University of Colorado Boulder, 
University of Oxford, University of Portsmouth, University of Utah, 
University of Virginia, University of Washington, University of Wisconsin, Vanderbilt University, and Yale University.


\bibliographystyle{mnras}
\bibliography{spin}



\appendix

\section{Two-fold ambiguity of rotation}
\label{app:method2}

\begin{figure}
\centering\includegraphics[width=0.9\columnwidth]{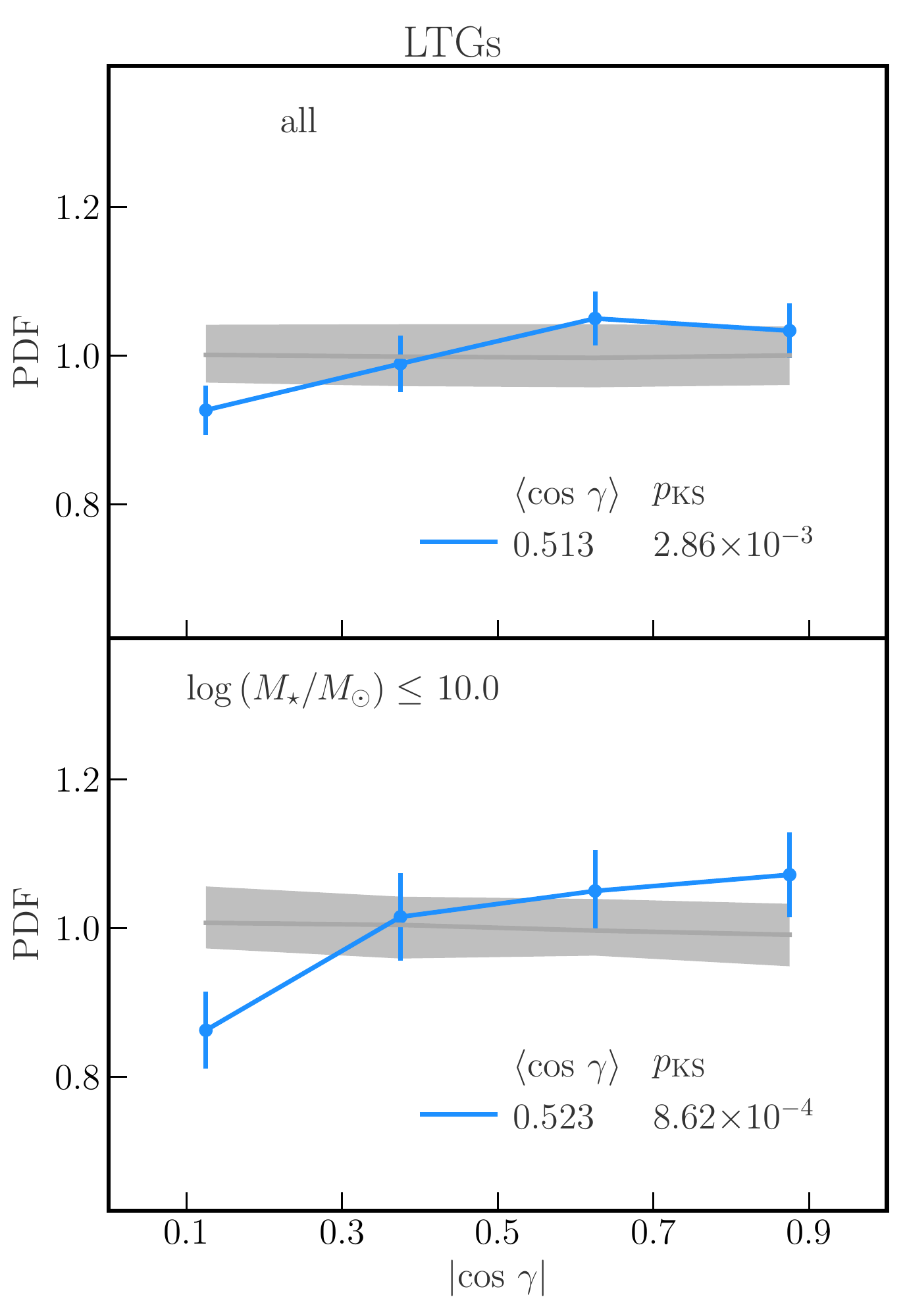}
\caption{Alignment between the filaments and spin of all ({\sl top}), and, low-mass ({\sl bottom}) LTGs. The error bars correspond to the bootstrap. The solid grey line and shaded area represent the median and 95 per cent confidence limits from 5000 random samples, respectively. The mean alignment angle together with the probability $p_{\rm KS}$ of the KS test are shown in each panel with corresponding symbols. 
LTGs in the probed mass range tend to have their spin parallel to their host filaments. The alignment signal seen for the entire population of LTGs is driven by the low mass sub-sample (\mstar $\leq 10^{10}$ \msun).
}
\label{fig:spin_ltgs_spin_pair}
\end{figure}

\begin{figure}
\centering\includegraphics[width=0.9\columnwidth]{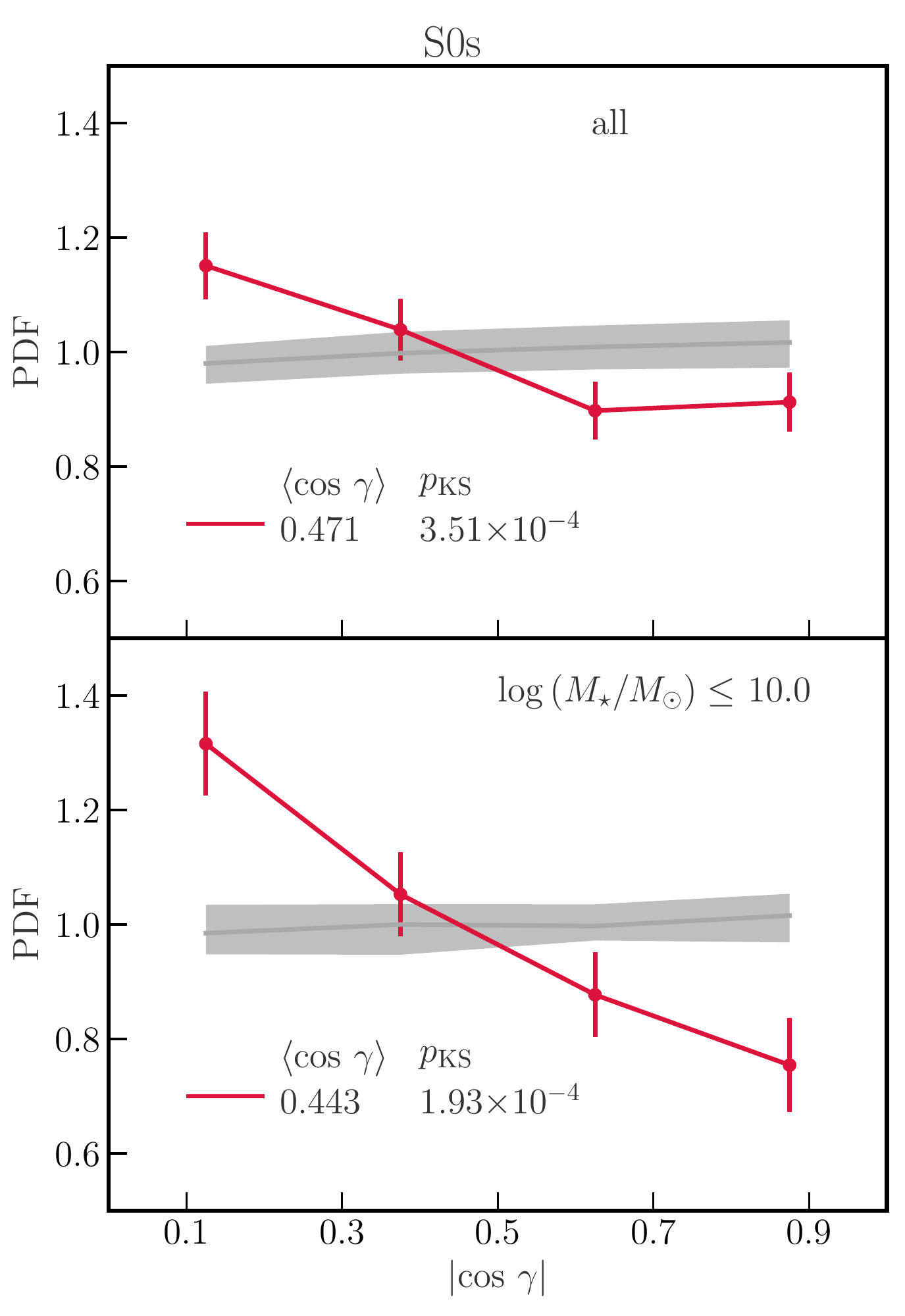}
\caption{Alignment between the filaments and spin of all ({\sl top}) and low-mass ({\sl bottom}) S0s.
The error bars correspond to the bootstrap.  The solid grey line and shaded area represent the median and 95 per cent confidence limits from 5000 random samples, respectively. The mean alignment angle together with the probability $p_{\rm {KS}}$ of the KS test are shown in each panel with corresponding symbols.
S0s in the probed mass range tend to have their spin perpendicular to their host filaments. 
The alignment signal seen for the entire population of S0s is driven by the low mass sub-sample (\mstar $\leq 10^{10}$ \msun). 
}
\label{fig:spin_S0s_spin_pair}
\end{figure}

Figures~\ref{fig:spin_ltgs_spin_pair} and \ref{fig:spin_S0s_spin_pair} show the measured alignment signal when statistically accounting for the clock-wise and counter-clockwise rotation, as proposed by
\cite{Lee2011}. This consists in assigning to each galaxy 
a set of two unit spin vectors differing from each other by the sign of $\hat{L}_r$, i.e. in addition to the spin vector defined by Eq.\ref{eq:spin}, we assign to each galaxy spin vector defined as 

\begin{equation}
\label{eq:spin_negative}
\begin{split}
    \hat{L}^{-}_x & = -\hat{L}_r \sin \alpha \cos \beta + \hat{L}_{\theta} \cos \alpha \cos \beta - \hat{L}_{\phi} \sin \beta \\
    \hat{L}^{-}_y & = -\hat{L}_r \sin \alpha \sin \beta + \hat{L}_{\theta} \cos \alpha \sin \beta + \hat{L}_{\phi} \cos \beta \\
    \hat{L}^{-}_z & = -\hat{L}_r \cos \alpha - \hat{L}_{\theta} \sin \alpha.
\end{split}    
\end{equation}

LTGs (Fig.~\ref{fig:spin_ltgs_spin_pair}) tend to have their spin parallel to their host closest filaments, while S0s (Fig.~\ref{fig:spin_S0s_spin_pair}) tend to have their spin in the orthogonal direction.

\bsp	
\label{lastpage}
\end{document}